\documentclass[letterpaper, 10 pt, conference]{ieeeconf}  

\IEEEoverridecommandlockouts                              
\usepackage[T1]{fontenc}

\usepackage{nccmath}

\usepackage{amssymb}    
\usepackage{mathtools}  
\usepackage{pgfplots}
\usepackage{tikz}
\usetikzlibrary{shapes}
\usetikzlibrary{positioning}
\usetikzlibrary{arrows,decorations.markings}
\usepackage{bm}

\usepackage{enumerate}

\usepackage{caption}
\usepackage{subcaption}
\usepackage{siunitx}

\usepackage{pgfplots}
\pgfplotsset{compat=1.18}
\usepackage{tikz}
\usepackage[hidelinks]{hyperref}
\usetikzlibrary{shapes}
\usetikzlibrary{positioning}
\usetikzlibrary{arrows,decorations.markings}
\usetikzlibrary{spy}

\newtheorem{theorem}{Theorem}
\newtheorem{assumption}{Assumption}

\newtheorem{definition}{Definition}
\newtheorem{lemma}{Lemma}

\newtheorem{remark}{Remark}

\title{\LARGE \bf
Event-triggered Robust Model Predictive Control\\under Hard Computation Resource Constraints
}

\author{Alexander Gräfe and Sebastian Trimpe
\thanks{This work was supported by the German Research Foundation (DFG) within the priority program 1914 (grant TR 1433/2). The authors gratefully acknowledge the computing time provided to them at the NHR Center NHR4CES at RWTH Aachen University (project number p0022034). This is funded by the Federal Ministry of Education and Research, and the state governments participating on the basis of the resolutions of the GWK for national high performance computing at universities.}%
\thanks{RWTHgpt (GPT4) has been used to assist with language editing.}
\thanks{Alexander Gräfe and Sebastian Trimpe are with the Institute for Data Science in Mechanical Enginnering, Faculty of Mechanical Engineering, RWTH Aachen University, 52068 Aachen, Germany
{\tt\small alexander.graefe@dsme.rwth-aachen.de, trimpe@dsme.rwth-aachen.de}}%
}

\begin{document}

{\onecolumn \begin{center} Accepted for publication at the proceedings of the 23rd European Control Conference\end{center}
		
		\noindent\fbox{%
			\parbox{\textwidth}{%
				© 2025 IEEE. Personal use of this material is permitted. Permission from IEEE must be obtained for all other uses, in any current or future media, including reprinting/republishing this material for advertising or promotional purposes, creating new collective works, for resale or redistribution to servers or lists, or reuse of any copyrighted component of this work in other works.
			}%
		}
	}
\twocolumn
		\newpage

\maketitle
\thispagestyle{empty}
\pagestyle{empty}

\begin{abstract}


Model predictive control (MPC) is capable of controlling nonlinear systems with guaranteed constraint satisfaction and stability.
However, MPC requires solving optimization problems online periodically, which often exceeds the local system's computational capabilities. 
A potential solution is to leverage external processing, such as a central industrial server.
Yet, this central computer typically serves multiple systems simultaneously, leading to significant hardware demands due to the need to solve numerous optimization problems concurrently.
In this work, we tackle this challenge by developing an event-triggered model predictive control (ET-MPC) that provably stabilizes multiple nonlinear systems under disturbances while solving only optimization problems for a fixed-size subset at any given time. 
Unlike existing ET-MPC methods, which primarily reduce average computational load yet still require hardware capable of handling all systems simultaneously, our approach reduces the worst-case computational load.
This significantly lowers central server hardware requirements by diminishing peak computational demands.
We achieve our improvements by leveraging recent advancements in distributed event-triggered linear control and integrating them with a robust MPC that employs constraint tightening.

\end{abstract}

	\newcommand{\fakepar}[1]{\vspace{1mm}\noindent\textbf{#1.}}
\newcommand{\capt}[1]{\mdseries{\emph{#1}}}
\newcommand*{\matlab}{\textit{Matlab}}
\newcommand*{\simulink}{\textit{Simulink}}
\newcommand*{\purepursuit}{\textit{Pure Pursuit}}
\newcommand*{\carCup}[0]{\textit{Carolo-Cup}~}
\newcommand*{\ros}[0]{\textit{ROS2}~}
\newcommand*{\psaf}[0]{Projektseminar Autonomes Fahren~}
\newcommand*{\darpa}[0]{\textit{DARPA}~}
\newcommand*{\opencv}[1]{\textit{OpenCV}}

\newcommand*{\jt}{\textit{Jetson TX2}}
\newcommand*{\jpi}{\textit{SDK Manager}}

\newcommand*{\phil}{\ensuremath{\varphi_\textrm{L}}}
\newcommand*{\philmax}{\ensuremath{\varphi_\textrm{L,max}}}
\newcommand*{\dphilmax}{\ensuremath{\dot{\varphi_\textrm{L,max}}}}

\newcommand*{\mat}[1]{{\ensuremath{\mathrm{\textbf{#1}}}}}
\newcommand*{\ma}[1]{{\ensuremath{\boldsymbol{\mathrm{#1}}}}}
\newcommand*{\mas}[1]{\ensuremath{\boldsymbol{#1}}}
\newcommand*{\ve}[1]{{\ensuremath{\boldsymbol{#1}}}}
\newcommand*{\ves}[1]{\ensuremath{\boldsymbol{\mathrm{#1}}}}

\newcommand*{\AP}{\ensuremath{\mathrm{AP}}}
\newcommand*{\doti}{\ensuremath{(i)^\cdot}}

\newcommand*{\inprod}[2]{\ensuremath{\langle #1,\,#2 \rangle}}


\newcommand*{\ud}{\ensuremath{\mathrm{d}}}

\newcommand*{\tn}[1]{\textnormal{#1}}

\newcommand*{\mrm}[1]{\ensuremath{\mathrm{#1}}}

\newcommand*{\transp}{\ensuremath{\mathrm{T}}}

\newcommand*{\rang}{\ensuremath{\operatorname{rg}}}

\newcommand*{\grpsb}[2]{\ensuremath{\left(#1\right)_{#2}}}
\newcommand*{\grprsb}[2]{\ensuremath{\left(#1\right)_{\mathrm{#2}}}}

\newcommand*{\normd}[2]{\ensuremath{\frac{\mathrm{d}#1}{\mathrm{d}#2}}}
\newcommand*{\normdat}[3]{\ensuremath{\left.\frac{\mathrm{d} #1}{\mathrm{d} #2}\right|_{#3}}}

\newcommand*{\matd}[2]{\ensuremath{\frac{\mathrm{D} #1}{\mathrm{D} #2}}}
\newcommand*{\matdat}[3]{\ensuremath{\left.\frac{\mathrm{D} #1}{\mathrm{D} #2}\right|_{#3}}}

\newcommand*{\partiald}[2]{\ensuremath{\frac{\partial #1}{\partial #2}}}
\newcommand*{\partialdat}[3]{\ensuremath{\left.\frac{\partial #1}{\partial #2}\right|_{#3}}}

\newcommand*{\FT}[1]{\ensuremath{\mathfrak{F}\left\{#1\right\}}}
\newcommand*{\FTabs}[1]{\ensuremath{\left|\mathfrak{F}\left\{#1\right\}\right|}}
\newcommand*{\IFT}[1]{\ensuremath{\mathfrak{F}^{-1}\left\{#1\right\}}}
\newcommand*{\DFT}[1]{\ensuremath{\mathrm{DFT}\left\{#1\right\}}}
\newcommand*{\DFTabs}[1]{\ensuremath{\left|\mathrm{DFT}\left\{#1\right\}\right|}}
\newcommand*{\Laplace}[1]{\ensuremath{\mathfrak{L}\left(#1\right)}}
\newcommand*{\InvLaplace}[1]{\ensuremath{\mathfrak{L^{-1}}\left(#1\right)}}
\newcommand*{\invtrans}{\ensuremath{\quad\bullet\!\!-\!\!\!-\!\!\circ\quad}}
\newcommand*{\trans}{\ensuremath{\quad\circ\!\!-\!\!\!-\!\!\bullet\quad}}

\newcommand*{\textcompstdfont}[1]{{\fontfamily{cmr} \fontseries{m} \fontshape{n} \selectfont #1}}

\newcommand*{\mlfct}[1]{\texttt{#1}}

\newcommand*{\UL}[2]{#1_\mathrm{#2}}
\newcommand*{\ULi}[2]{#1_{#2}}
\newcommand*{\dy}[0]{\dot{y}}
\newcommand*{\ddy}[0]{\ddot{y}}
\newcommand*{\receivedPower}[0]{\frac{A}{y^2+h^2}}
\newcommand*{\dD}[0]{\dot{D}}
\newcommand*{\tf}[0]{\UL{t}{f}}
\newcommand*{\umax}[0]{\UL{u}{max}}

\newcommand*{\x}[1]{\UL{x}{#1}}
\newcommand*{\xv}[0]{\ve{x}}
\newcommand*{\la}[1]{\UL{\lambda}{#1}}
\newcommand*{\ua}[0]{\UL{u}{+}}
\newcommand*{\ub}[0]{\UL{u}{-}}
\newcommand*{\Ht}[1]{\tilde{H}(#1)}
\newcommand*{\Dp}[0]{\UL{D}{p}}
\newcommand*{\Dm}[0]{\overline{\dot{D}}}
\newcommand*{\fdynamic}[0]{\UL{\ve{f}}{dynamic}(\ve{y}, u)}
\newcommand*{\fE}[0]{\UL{f}{E}(\ve{y}, u)}
\newcommand*{\fD}[0]{\UL{f}{D}(\ve{y}, u)}
\newcommand*{\vl}[0]{\UL{v}{l}}
\newcommand{\vex}[0]{\ve{x}}
\renewcommand*{\of}[1]{(#1)}
\newcommand*{\vexof}[1]{\vex\of{#1}}
\newcommand*{\vexul}[1]{\UL{\vex}{#1}}
\newcommand*{\vexulof}[2]{\UL{\vex}{#1}\of{#2}}
\newcommand*{\vexuli}[1]{\ULi{\vex}{#1}}
\newcommand*{\hatvexuli}[1]{\ULi{\hat{\vex}}{#1}}
\newcommand*{\vexuliof}[2]{\ULi{\vex}{#1}\of{#2}}
\newcommand*{\hatvexuliof}[2]{\ULi{\hat{\vex}}{#1}\of{#2}}
\newcommand*{\pcom}[0]{{P\ul{com}}}
\newcommand*{\pcomhat}[0]{\hat{P}\ul{com}{}}
\newcommand{\weight}[2]{w\uliof{#1}{#2}}
\newcommand{\history}[0]{\mathcal{E}}
\newcommand{\weighthat}[2]{\hat{w}\uliof{#1}{#2}}

\newcommand*{\veerr}[0]{\ve{e}}
\renewcommand*{\ul}[1]{_{\mathrm{#1}} }
\newcommand*{\uliof}[2]{_{#1}{\of{#2}}}
\newcommand*{\uli}[1]{_{#1} }
\newcommand*{\koop}[0]{\mathcal{K}}
\newcommand*{\koopt}[0]{\mathcal{K}^t}
\newcommand*{\gof}[1]{g\of{#1}}
\newcommand*{\gul}[1]{\ULi{g}{#1}}
\newcommand*{\gofxof}[1]{\gof{\vexof{#1}}}
\newcommand*{\phivexof}[1]{\phi\of{\vexof{#1}}}
\newcommand*{\veg}[0]{\ve{g}}
\newcommand*{\vegof}[1]{\ve{g}\of{#1}}
\newcommand*{\vegulof}[2]{\ve{g}_{\mathrm{#1}}\of{#2}}
\newcommand*{\vey}[0]{\ve{y}}
\newcommand*{\veyof}[1]{\vey\of{#1}}
\newcommand*{\veyul}[1]{\UL{\vey}{#1}}
\newcommand*{\tzero}[0]{\UL{t}{0}}
\newcommand*{\vepsi}[0]{\ve{\psi}}
\newcommand*{\vephi}[0]{\ve{\phi}}
\newcommand*{\veu}[0]{\ve{u}}
\newcommand*{\veuof}[1]{\veu\of{#1}}
\newcommand*{\veuul}[1]{\UL{\veu}{#1}}
\newcommand*{\vev}[0]{\ve{v}}
\newcommand*{\vevof}[1]{\vev\of{#1}}
\newcommand*{\vevul}[1]{\UL{\vev}{#1}}
\newcommand*{\xul}[1]{\ULi{x}{#1}}
\newcommand*{\aul}[1]{\UL{a}{#1}}
\newcommand*{\kul}[1]{\UL{k}{#1}}
\newcommand*{\uul}[1]{\ULi{u}{#1}}

\newcommand*{\maAuli}[1]{\ULi{\ma{A}}{#1}}
\newcommand*{\veeps}[0]{\ve{\epsilon}}
\newcommand*{\twonorm}[1]{||#1||_\mathrm{2}}

\newcommand*{\lambdamaxof}[1]{\UL{\lambda}{max}\of{#1}}
\newcommand*{\lambdaminof}[1]{\UL{\lambda}{min}\of{#1}}
\newcommand*{\B}[0]{Method B\:}

\newcommand*{\tensorflowprob}{\textit{Tensorflow-Probability}~}

\newcommand*{\EX}[0]{\mathbb{E}}
\newcommand*{\tr}[0]{\mathrm{Trace}}
\newcommand*{\var }[0]{\mathrm{Var}}
\newcommand*{\given}[0]{\;\middle|\;}
\newcommand*{\gpess}[0]{g\ul{pess}{}}
\newcommand*{\gopt}[0]{g\ul{opt}{}}
\newcommand*{\tveerr}[0]{\tilde{\veerr}}
\newcommand*{\ofs}[1]{\left[#1\right]}

\newcommand*{\matA}[1]{\ma{A}\uli{#1}}
\newcommand*{\matGamma}[0]{\ma{\Gamma}}
\newcommand*{\matGammaOpt}[0]{\ma{\Gamma}\ul{opt}}

\newcommand*{\matQ}[1]{{\ma{Q}\uli{#1}}}
\newcommand*{\matP}[1]{\ma{P}\uli{#1}}
\newcommand*{\eps}[2]{\epsilon\uli{#1}\of{#2}}
\newcommand*{\alp}[2]{\alpha\uli{#1}\of{#2}}
\newcommand*{\pfail}[0]{P\ul{loss}}
\newcommand{\priority}[2]{g\uli{#1}\of{\errorest{#1#1}{#2}}}
\newcommand*{\diag}[0]{\text{diag}}
\newcommand*{\tp}[1]{\of{#1}}
\newcommand*{\state}[2]{\vex\uli{#1}\tp{#2}}
\newcommand*{\stateest}[2]{\hat{\vex}\uli{#1}\tp{#2}}
\newcommand*{\noise}[2]{\ve{w}\uli{#1}\tp{#2}}
\newcommand*{\processnoise}[2]{\ve{v}\uli{#1}\tp{#2}}
\newcommand*{\errart}[2]{\ve{\tilde{e}}\uli{#1}\tp{#2}}
\newcommand*{\errorest}[2]{\ve{e}\uli{#1}\tp{#2}}
\newcommand*{\barerrorest}[2]{\ve{\bar{e}}\uli{#1}\tp{#2}}
\newcommand*{\errorestmeas}[2]{\ve{e}\uli{#1, meas}\tp{#2}}
\newcommand*{\erroragents}[2]{\hat{\ve{e}}\uli{#1}\tp{#2}}
\newcommand*{\erroragentstotal}[2]{\underline{\hat{\ve{e}}}\uli{#1}\tp{#2}}
\newcommand*{\erroresttotal}[2]{\underline{\ve{e}}\uliof{#1}{#2}}
\newcommand*{\matB}[1]{\ma{B}\uli{#1}}
\newcommand*{\matV}[1]{\ma{V}\uli{#1}}
\newcommand*{\matVti}[1]{\ma{\tilde{V}}\uli{#1}}
\newcommand*{\matW}[1]{\ma{W}\uli{#1}}
\newcommand*{\matAT}[1]{\left(\ma{A}\uli{#1}^\transp\right)}
\newcommand*{\matF}[1]{\ma{F}\uli{#1}}
\newcommand*{\rec}[2]{\gamma\uli{#1}\tp{#2}}
\newcommand*{\Cov}{\mathrm{Cov}}
\newcommand*{\matAti}[1]{\tilde{\ma{A}}\uli{#1}}

\newcommand{\io}[1]{\Gamma\of{#1}}
\newcommand{\oi}[1]{\Lambda\of{#1}}

\newcommand{\matAd}[1]{\tilde{\mat{A}}\ul{d, \mathrm{#1}}}
\newcommand{\ri}[1]{{r\uli{#1}}}
\newcommand{\di}[1]{{d\uli{#1}}}
\newcommand{\nii}[1]{{n\uli{#1}}}
\newcommand{\deltai}[1]{\Delta\uli{#1}}
\newcommand{\betai}[1]{\beta\uli{#1}}
\newcommand{\alphai}[1]{\alpha\uli{#1}}
\newcommand{\zetai}[1]{\zeta\uli{#1}}
\newcommand{\lmax}[1]{\lambda\ul{max}\of{#1}}
\newcommand{\lmin}[1]{\lambda\ul{min}\of{#1}}
\newcommand{\unknown}[0]{\left(\cdot\right)}
\newcommand*{\kt}[0]{\tilde{k}}
\newcommand*{\nz}[0]{\ve{n}\ul{0}}
\newcommand*{\norm}[2]{||#1||_{#2}}
\newcommand*{\diff}[0]{\text{d}}

\newcommand{\dalpha}[0]{\delta\alpha}
\newcommand{\setS}[0]{\mathcal{S}}
\newcommand{\cone}[0]{c\ul{1}\of{\setS}}
\newcommand{\ctwo}[0]{c\ul{2}\of{\setS}}

\newcommand{\pos}[0]{\ve{p}}
\newcommand{\speed}[0]{\ve{v}}
\newcommand{\acc}[0]{\ve{a}}

\newcommand{\posset}[1]{\pos\uli{#1, \mathrm{set}}}
\newcommand{\speedset}[1]{\speed\uli{#1, \mathrm{set}}}
\newcommand{\accset}[1]{\acc\uli{#1, \mathrm{set}}}
\newcommand{\vexset}[1]{\vex\uli{#1, \mathrm{set}}}
\renewcommand{\posset}[1]{\pos\uli{#1}}
\renewcommand{\speedset}[1]{\speed\uli{#1}}
\renewcommand{\accset}[1]{\acc\uli{#1}}
\renewcommand{\vexset}[1]{\vex\uli{#1}}
\newcommand{\stimeobj}[0]{T\ul{o}}
\newcommand{\stimes}[0]{T\ul{s}}
\newcommand{\stimecons}[0]{T\ul{b}}
\newcommand{\stimecol}[0]{T\ul{c}}
\newcommand{\dxmax}[2]{\Delta\vexset{#1}{}\ul{max}\of{#2}}
\newcommand{\dpmax}[2]{\Delta\posset{#1}{}\ul{,max}\of{#2}}

\newcommand{\cu}{CU}
\newcommand{\cus}{CUs}
\newcommand{\artinput}{\ve{u}}
\newcommand{\artinputset}{\mathcal{u}}

\newcommand{\gp}[2]{\mathcal{GP}\of{#1, #2}}
\newcommand{\kernel}[3]{k\ul{#1}\of{#2, #3}}
\newcommand{\tkernel}[2]{\kernel{t}{#1}{#2}}
\newcommand{\xkernel}[2]{\kernel{x}{#1}{#2}}
\newcommand{\mkernel}[3]{K\ul{#1}\of{#2, #3}}
\newcommand{\tmkernel}[2]{\mkernel{t}{#1}{#2}}
\newcommand{\xmkernel}[2]{\mkernel{x}{#1}{#2}}
\newcommand{\realnmbrs}{\mathbb{R}}

\newcommand{\tobjfunc}[3]{J\ul{#1}\of{[#2, #3]}}
\newcommand{\tobjfuncn}[3]{\tilde{J}\ul{#1}\of{[#2, #3]}}

\newcommand{\opttheta}[2]{\theta^*\uli{#1}\of{#2}}

\newcommand{\jmin}{J_\mathrm{min}}
\newcommand{\sx}{\sigma_\mathrm{x}}
\newcommand{\st}{\sigma_\mathrm{t}}

\newcommand{\taua}{\tau_\mathrm{a}}
\newcommand{\taub}{\tau_\mathrm{b}}

\newcommand{\xpredi}[3]{x_{#1}(#2|#3)}
\newcommand{\xpredioptim}[3]{\bar{x}_{#1}(#2|#3)}
\newcommand{\xpred}[2]{x(#1|#2)}
\newcommand{\xreali}[2]{x_{#1}(#2)}
\newcommand{\xreal}[1]{x(#1)}
\newcommand{\xpredopt}[2]{x^*(#1|#2)}
\newcommand{\xpredopti}[3]{x^*_{#1}(#2|#3)}
\newcommand{\xpredcand}[2]{\tilde{x}(#1|#2)}
\newcommand{\xpredcandi}[3]{\tilde{x}_{#1}(#2|#3)}

\newcommand{\wart}[2]{w_\mathrm{art}(#1|#2)}

\newcommand{\upredi}[3]{u_{#1}(#2|#3)}
\newcommand{\upredioptim}[3]{\bar{u}_{#1}(#2|#3)}
\newcommand{\upred}[2]{u(#1|#2)}
\newcommand{\ureali}[2]{u_{#1}(#2)}
\newcommand{\ureal}[1]{u(#1)}
\newcommand{\upredopt}[2]{u^*(#1|#2)}
\newcommand{\upredopti}[3]{u^*_{#1}(#2|#3)}
\newcommand{\upredcand}[2]{\tilde{u}(#1|#2)}
\newcommand{\upredcandi}[3]{\tilde{u}_{#1}(#2|#3)}

\newcommand{\lf}[0]{L_\mathrm{f}}
\newcommand{\tc}[0]{t_\mathrm{c}}
\newcommand{\tcn}[1]{t_\mathrm{#1}}

\newcommand{\vdelta}[0]{V_\mathrm{\delta}}

\newcommand{\cdl}[0]{c_\mathrm{\delta,\ell}}
\newcommand{\cdu}[0]{c_\mathrm{\delta,u}}
\newcommand{\deltaloc}[0]{\delta_\mathrm{loc}}
\newcommand{\kmax}[0]{k_\mathrm{max}}

\newcommand{\vf}[0]{V_\mathrm{f}}
\newcommand{\vw}[0]{\tilde{V}}
\newcommand{\vwmax}[0]{\tilde{V}_\mathrm{max}}

\newcommand{\xf}[0]{\mathcal{X}_\mathrm{f}}
\newcommand{\kappaf}[0]{\kappa_\mathrm{f}}

\newcommand{\rmq}[0]{\mathrm{Q}}
\newcommand{\rmr}[0]{\mathrm{R}}
\newcommand{\rmp}[0]{\mathrm{P}}
\newcommand{\vmax}[0]{V_\mathrm{max}}

\newcommand{\lipschitzo}[0]{L_{o}}
\newcommand{\lipschitzd}[0]{L_{d}}
\newcommand{\rmin}[0]{r_{\mathrm{min}}}

	\section{INTRODUCTION}

	Model predictive control (MPC) is a state-of-the-art technique to control non-linear systems with guaranteed constraint satisfaction and stability~\cite{rawlings2017model}.
	However, MPC needs to solve an optimization problem at every time step, demanding a lot of computation power that often surpasses the system's resources available locally.
	One approach to address this challenge is external processing, where the system transmits its state to a central computer, such as an industrial or cloud server, which solves the optimization problem and returns the control input~\cite{xia2022brief}.
	In this setup, the central computer usually serves multiple systems at once.
	However, this requires the central computer hardware to solve the optimization problems for all connected systems concurrently, which results in a high computational load.
	
	Event-triggered MPC (ET-MPC) addresses this challenge by only solving the optimization problem when a condition is met, e.g., when the distance between the actual and predicted state trajectory reaches a threshold~\cite{li2014event, brunner2017robust, liu2019codesign}. 
	This approach has shown a significant computational load reduction compared to traditional periodic MPCs.
	
\begin{figure}[t]
		\centering
		\includegraphics[width=0.9\linewidth]{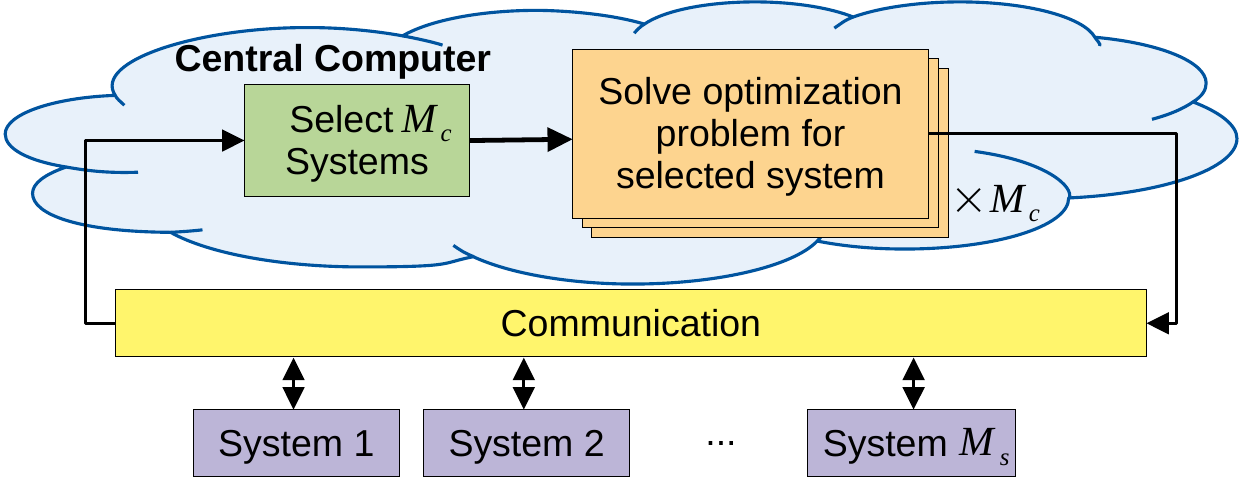}
		\vspace{-0.1cm}
		\caption{Overview of the system setup. \capt{A central computer controls $M_\mathrm{s}$ independent systems using MPC. Due to its limited resources, it can only handle $M_\mathrm{c} < M_\mathrm{s}$ MPC optimizations per time step. To address this limitation, an event-triggered approach is employed to prioritize and select the systems in most need (green box).}}
		\vspace{-0.6cm}
		\label{fig:overview}
	\end{figure}

	However, most approaches only save computation load \emph{on average} and not in the \emph{worst-case}.
	Consequently, the central computer server hardware must still be capable of solving the optimization problems for all systems at any given moment.
	Most methods thus do not reduce the necessary hardware or the associated expenses of the central computer.
	In this work, we thus ask (Fig.~\ref{fig:overview}): \emph{How can we control $M_\mathrm{s}$ non-linear systems using MPC, when the central computer can only serve $M_\mathrm{c}<M_\mathrm{s}$ of them at once?}
	
	So far, only two approaches tackling ET-MPC reducing worst-case load exist.
	In a previous work~\cite{grafe2022event}, we developed an ET-MPC for drone swarms that ensures collision avoidance while solving optimization problems for only a maximum number of drones simultaneously.
	However, this ET-MPC is tailored to its application, limited to linear systems without disturbances and lacks stability guarantees.
	The second approach by Chen et al.~\cite{chen2024reconfigurable} controls multiple systems with coupled constraints, reducing the worst-case load.
	However, it only focuses on undisturbed linear systems and features no feasibility or stability guarantees.

	In contrast, this work focuses on independent non-linear systems with disturbances. 
	Our robust ET-MPC achieves constraint satisfaction and stability while only solving an optimization problem for $M_\mathrm{c}<M_\mathrm{s}$ systems at every time (Fig.~\ref{fig:overview}).
	We combine proof techniques for event-triggers under resource constraints~\cite{mager2022scaling} with constraint tightening for robust MPC~\cite{kohler2018novel}.

	\textbf{Notation.} We note $||\cdot||$ the vector/matrix-two-norm and define $||x||_A = \sqrt{x^TAx}$. We call a function $g: \mathbb{R}_{0}^+\to\mathbb{R}_{0}^+$, which is strictly increasing and for which $g(0)=0$, a $\mathcal{K}$-function and a $\mathcal{K}_\infty$ function if $\lim_{x\to\infty}g(x)=\infty$. A $h: \mathbb{R}_{0}^+\times\mathbb{R}_{0}^+\to\mathbb{R}_{0}^+$ is called $\mathcal{KL}$-function, if it is a $\mathcal{K}_\infty$ function in the first argument and if it converges to zero when the second argument goes to infinity~\cite{jiang2001input}. The value at time $t+k$ of a trajectory $a$ starting at time $t$ is~$a(k|t)$.

	\section{PROBLEM SETTING}

	We have $M_\mathrm{s}$ systems with independent discrete-time dynamics for all $t\in\mathbb{N}_0$ and $i\in\{1, \cdots , M_\mathrm{s} \}$ 
	\begin{align}
		\label{eq:system}
		\xreali{i}{t+1} = f(\xreali{i}{t},\ureali{i}{t}) + w_i(t),\\
		\xreali{i}{0} = x_{i, 0},\text{~}||w_i(t)|| \leq \hat{w}.\nonumber
	\end{align}
	Each system has a state $\xreali{i}{t}\in\mathbb{R}^n$, an input $\ureali{i}{t}\in\mathbb{R}^m$, an equilibrium at $x_{i}=0,u_i=0$, an initial state $x_{i, 0}$ and is disturbed by $w_i(t)\in\mathbb{R}^n$, whose norm is bounded by $\hat{w}$.
	The goal is to find inputs $\ureali{i}{t}$, such that the systems are stable and fulfill compact state and input constraints: $\forall t\in\mathbb{N}_0:$ $x_i\in\mathcal{X} := \{x \in\mathbb{R}^n:Hx\leq\mathbf{1}_\mathrm{H}\}$ and $u_i\in\mathcal{U}:= \{u \in\mathbb{R}^m:Lu\leq\mathbf{1}_\mathrm{L}\}$, where $\mathbf{1}$ is a vector full of ones.
	We consider a generalized version of input-to-state stability (ISS)~\cite{jiang2001input}, called gISS, which helps to simplify our derivations.
	
	\begin{definition}[gISS] A system $x(t+1) = h(x(t)) + w(t)$ with $||w(t)||\leq \hat{w}$ is called gISS on set $\mathcal{X}_0\subseteq\mathbb{R}^n$, if there exists a function $\omega: \mathbb{R}_0^+\times\mathbb{R}_0^+\times\mathbb{N}_0\to\mathbb{R}_0^+$ that satisfies:
    \begin{enumerate}[(i)]
    \item $\forall t \in \mathbb{N}_0 $, $\omega(0, 0, t) = 0$,
    \item $\forall t \in \mathbb{N}_0 $, $\forall a_1, a_2, b \in\mathbb{R}_0^+ $ with $ a_1 < a_2 $: $\omega(a_1, b, t) < \omega(a_2, b, t)$ and $\forall a, b_1, b_2 \in\mathbb{R}_0^+ $ with $ b_1 < b_2 $: $\omega(a, b_1, t) < \omega(a, b_2, t)$,
    \item $\forall a\in\mathbb{R}_0^+: \lim_{t\to\infty}\omega(a, \hat{w}, t)=\phi(\hat{w})$, where $\phi:\mathbb{R}_0^+\to\mathbb{R}_0^+$ is a $\mathcal{K}$-Function,
    \item $\forall t\in \mathbb{N}_0, \forall x(0)\in\mathcal{X}_0:||x(t)|| \leq \omega(||x(0)||, \hat{w}, t)$.
\end{enumerate}
\end{definition}
	\begin{remark}
		ISS considers two summands in the last property, one converging to zero for $t\to\infty$ and the second being constantly growing with $\hat{w}$.
		gISS has the same behavior.
		The state's norm converges to a constant regime $\phi(\hat{w})$ for $t\to\infty$ that constantly grows with $\hat{w}$.
	\end{remark}

	The setting features a central computer, which can calculate control trajectories $\psi(k|t, \xreali{i}{t})$ for $M_\mathrm{c}$ systems per timestep $t$ at once using optimization.
	At each time $t$, it hence must select $M_\mathrm{c}$ out of the $M_\mathrm{s}$ systems.
	We call the algorithm making this decision event-trigger (ET) in the following.

	The other 
	systems reuse their old control trajectory.
	Formally, the control trajectory of a system $i$ at time $t$ is

	\begin{equation}
		\upredi{i}{k}{t} = \begin{cases}
			\upredi{i}{k+1}{t-1} & \text{if no recalculation} \\
			\psi(k|t, \xreali{i}{t}) & \text{else}
		\end{cases},
	\end{equation}
	where we initialize $\upredi{i}{\cdot}{0} = 0$.
	The system then applies the first part of this trajectory on its input $\ureali{i}{t} = \upredi{i}{0}{t}$.

	In summary, per timestep $t$, $M_\mathrm{c}$ systems apply an input derived from their current state $\xreali{i}{t}$.
	The remaining $M_\mathrm{s} - M_\mathrm{c}$ systems apply inputs based on a past state.
	We demonstrate how to effectively control this configuration to ensure gISS and constraint satisfaction across all systems. 
	To achieve this, we must address two key questions.
	\emph{How does the central computer choose the systems it recalculates? And, how does the computer calculate $\psi$?}
	We answer the first question in Section~\ref{sec:trigger} and the second in Section~\ref{sec:mpc}, making common additional assumptions on system (\ref{eq:system}) along the way.

	\section{EVENT-TRIGGER FOR RESOURCE CONSTRAINED ROBUST MPC}
	\label{sec:trigger}

	We use a priority based event-trigger~\cite{grafe2022event,mager2022scaling}. 
	At every time, the central computer calculates priorities $g_i(t)$  measuring the need for recomputation of system $i$.
	For this, we define the predicted state trajectory as
	\begin{align}
		\xpredi{i}{k+1}{t} &= f(\xpredi{i}{k}{t},\upredi{i}{k}{t})\\
		\xpredi{i}{0}{t} &= \begin{cases}
			\xpredi{i}{1}{t-1} & \text{if no recalculation} \\
			\xreali{i}{t} & \text{else}
		\end{cases}.
	\end{align}
	The priorites are 
	\begin{equation}
	g_i(t)=||\xreali{i}{t} - \xpredi{i}{1}{t-1}||.
	\end{equation}
	The central computer selects the systems with the $M_\mathrm{c}$-highest priorities for computation.
	For the first $t<\lceil\frac{M_\mathrm{s}}{M_\mathrm{c}}\rceil$, it select the systems using round-robin instead, such that all systems are recalculated at least once. We now prove that this ET leads to a bounded distance between predicted and real state.
	
	\fakepar{Assumption} For this, we make the following assumption.

	\begin{assumption}
    \label{as:lipschitz}
    Define the non-disturbed system as
    \begin{equation}
        z\uli{i}(t+1) = f(z\uli{i}(t),\ureali{i}{t}).
    \end{equation}
    There exists a function $\vw: \mathbb{R}_0^+\times\mathbb{R}_0^+\times\mathbb{N}_0\to\mathbb{R}$ such that 
    \begin{enumerate}[(i)]
        \item $\forall t\in\mathbb{N}_0:\vw(0, 0, t) = 0$,
        \item $\forall a_1, a_2, b \in\mathbb{R}_0^+, c\in\mathbb{N}_0 $ with $ a_1 < a_2 $: $\vw(a_1, b, c) < \vw(a_2, b, c)$; $\forall a, b_1, b_2 \in\mathbb{R}_0^+, c\in\mathbb{N}_0 $ with $ b_1 < b_2 $: $\vw(a, b_1, c) < \vw(a, b_2, c)$; and $\forall a, b \in\mathbb{R}_0^+, c_1,c_2\in\mathbb{N}_0 $ with $ c_1 < c_2 $: $\vw(a, b, c_1) < \vw(a, b, c_2)$,
    \item $t, \tau\in\mathbb{N}: ||\xreali{i}{t+\tau} - z\uli{i}(t+\tau)|| \leq \vw(\hat{w},||\xreali{i}{t} - z\uli{i}(t)||, \tau)$.
    \end{enumerate}
\end{assumption}
	\vspace{0.5em}
	This assumption gives a bound on the error propagation of state predictions. 
	If $f$ is Lipschitz continuous in the first argument with constant $L$, a possible choice is $\vw(\hat{w},\Delta,\tau) = L^\tau\Delta+(L^\tau-1)/(L-1)\hat{w}$.
	But, tighter bounds might exist, e.g., when the system is prestabilized with a local controller.

	\fakepar{Theoretical analysis}
	We now show that this trigger leads to a bounded difference between real and predicted state.

	\begin{lemma}
    \label{lem:boundedstatediff}
    If Assumption \ref{as:lipschitz} holds, then
    $\forall\hat{w} \in \mathbb{R}^+_0$, $\forall t\in\mathbb{N}_0$ with $p = \lceil\frac{M_\mathrm{s}}{M_\mathrm{c}}\rceil$
    \begin{align}
        \label{eq:vwmax}
        ||\xreali{i}{t} - \xpredi{i}{1}{t-1}|| \leq& \max_{\tau\in[0, p]}\vw(\hat{w},\vw(\hat{w},0, p-\tau), \tau)\nonumber\\
        =:&\vwmax(\hat{w}).
    \end{align}
\end{lemma}
	
	\begin{proof}
		\textbf{Step 1:} We know that when the agent was recalculated at timestep $t_\mathrm{c}$, then for all $t > t_\mathrm{c}$
		\begin{align}
			||\xreali{i}{t} - \xpredi{i}{1}{t-1}|| =& ||\xreali{i}{t} - \xpredi{i}{t-\tc}{\tc}|| \nonumber\\\leq& \vw(\hat{w},0,t-\tc).
		\end{align}
		\textbf{Step 2:}
		We know that for $t=p$, the agent was once recalculated since $t=0$ at $t_\mathrm{c0}\in[0, p-1]$. Hence 
		\begin{align}
			||\xreali{i}{p} - &\xpredi{i}{1}{p-1}|| = ||\xreali{i}{t} - \xpredi{i}{t-\tcn{0}}{\tcn{0}}|| \nonumber\\ \leq& \vw(\hat{w},0,p-t_\mathrm{c0}) \leq \vw(\hat{w},0,p)\leq\vwmax.
		\end{align}
		\textbf{Step 3:} For times $t>p$, we look at the interval $[t-p, t]$. There exist two cases~\cite{mager2022scaling}.

		\noindent\textit{Case 1.} The system $i$ was recalculated at time $\tcn{1}\in [t-p, t-1]$. Then 
		\begin{align}
			||\xreali{i}{t} - &\xpredi{i}{1}{t-1}|| = ||\xreali{i}{t} - \xpredi{i}{t-\tcn{1}}{\tcn{1}}|| \nonumber\\
			\leq&  \vw(\hat{w},0,p-\tcn{1}) \leq \vw(\hat{w},0,p)\leq\vwmax(\hat{w}) .
		\end{align}
		\noindent\textit{Case 2.} The system was not recalculated in $[t-p, t-1]$. Because $p = \lceil\frac{M_\mathrm{s}}{M_\mathrm{c}}\rceil$, there was enough time in the interval $[t-p,t-1]$ for every system to be recalculated at least once. If system $i$ was not recalculated, then there exists a system $j$, which was recalculated twice at times $\tcn{1} \geq t-p$ and $\tcn{2}$ with $t-1 \geq \tcn{2} > \tcn{1}$. At the second time $\tcn{2}$, it is
		\begin{align}
			g_i(\tcn{2}) = ||\xreali{i}{\tcn{2}} -& \xpredi{i}{1}{\tcn{2}-1}|| \nonumber \\ 
			\leq g_j(\tcn{2}) = &||\xreali{j}{\tcn{2}} - \xpredi{j}{1}{\tcn{2}-1}||\nonumber\\
				\leq & \vw(\hat{w}, 0, \tcn{2} - \tcn{1}),
		\end{align}
		because of the trigger and hence
		\begin{align}
			||\xreali{i}{t} -& \xpredi{i}{1}{t-1}||\nonumber \\
			\leq& \vw(\hat{w}, \vw(\hat{w}, 0, \tcn{2} - \tcn{1}), t - \tcn{2})\nonumber\\
			\leq&\vw(\hat{w}, \vw(\hat{w}, 0, \tcn{2} - (t-p)), t - \tcn{2})\nonumber\\
			\leq& \max_{\tau\in[0, p]}\vw(\hat{w}, \vw(\hat{w}, 0, p-\tau), \tau)
		\end{align}
	\end{proof}
	For the next proofs, we need the following property.
	\begin{lemma}
		\label{lem:vwmaxkfunction}
		$\vwmax(\hat{w})$ is a $\mathcal{K}$-function.
	\end{lemma}
	\begin{proof}
		It is $\forall \tau\in\mathbb{N}_0:\vw(0,\vw(0,0, p-\tau), \tau)=0$ and hence $\vwmax(0)=0$.
		For all $\hat{w}_1<\hat{w}_2$ with $\tau_1=\arg \max_{\tau\in[0, p]}\vw(\hat{w}_1,\vw(\hat{w}_1,0, p-\tau), \tau)$, it holds $\vwmax(\hat{w}_1)=\vw(\hat{w}_1,\vw(\hat{w}_1,0, p-\tau_1), \tau_1)<\vw(\hat{w}_2,\vw(\hat{w}_2,0, p-\tau_1), \tau_1)\leq\vwmax(\hat{w}_2)$.
	\end{proof}

\section{ROBUST MODEL PREDICTIVE CONTROL}
\label{sec:mpc}
	Based on this ET, we design an MPC, i.e., how the central computer calculates $\psi$.
	The intuition behind our design stems from Lemma~\ref{lem:boundedstatediff}.
	Treating the difference between predicted state and the real state as a bounded disturbance, we leverage an MPC robust against such disturbances. In the following, we drop the index $i$ of the system for brevity. 
	
	The central computer solves the following optimization problem for each selected system:
	\begin{align}
		V_n(x(t)) =& \min_{\upred{\cdot}{t}}\sum_{k=0}^{N-1}l(\xpred{k}{t}, \upred{k}{t}) + ||\xpred{N}{t}||^2_\mathrm{P}\nonumber\\
	\text{s.t.~}\forall k\in\{0, 1 \cdots,& N-1\}:\nonumber\\
	\xpred{k+1}{t}=&f(\xpred{k}{t},\upred{k}{t}, \text{~}\xpred{0}{t} = \xreal{t},\nonumber\\
	\xpred{k}{t}\in&\tilde{\mathcal{X}}_k, \text{~}\upred{k}{t}\in\tilde{\mathcal{U}}_k,
	\xpred{N}{t}\in\xf,\label{eq:optimizationproblem}
	\end{align}
	where $l(\xpred{k}{t}, \upred{k}{t}) = ||\xpred{k}{t}||^2_\mathrm{Q} + ||\upred{k}{t}||^2_\mathrm{R})$, $Q, R, P\succ 0$ and $\xf$ as final state constraint.
	We use constraint tightening with decay rate $\rho$ , a tunable factor $\epsilon\in\mathbb{R}^+$~\cite{kohler2018novel} and $\epsilon_k = (1-\sqrt{\rho}^k) / (1-\sqrt{\rho})\epsilon$
	\begin{align}
		\tilde{\mathcal{X}}_k = \{x\in\mathbb{R}^n | H x\leq (1 - \epsilon_k)\mathbf{1}_\mathrm{H}\}\\
		\tilde{\mathcal{U}}_k = \{u\in\mathbb{R}^m | L x\leq (1-\epsilon_k)\mathbf{1}_\mathrm{L}\}.
	\end{align}
	The optimization problem yields a trajectory for $k\leq N$.
	For $k>N$, we use the controller $\kappaf$ of the following Assumption~\ref{as:localcontroller} to predict states and control inputs.

	\fakepar{Assumptions} We now derive constraint satisfaction and gISS of the proposed ET-MPC.
	We first need the following assumptions from robust MPC~\cite{kohler2018novel, hertneck2018learning}.

	\begin{assumption}[Local incremental stability]
    \label{as:incrementalstability}
    
    There exists a control law $\kappa:\mathcal{X}\times\mathcal{X}\times\mathcal{U}\to\mathbb{R}^m$, a $\delta$-Lyapunov-function $\vdelta:\mathcal{X}\times\mathcal{X}\to\mathbb{R}_{\geq 0}$ that is continuous in the first argument and $\vdelta(x,x)=0, \forall x\in\mathcal{X}$ and parameters $\cdl$, $\cdu$, $\deltaloc$ and $\kmax$ $\in\mathbb{R}_{\geq0}$, $\rho\in(0, 1)$ such that the following properties hold for all $x, z\in\mathcal{X}$ with $\vdelta(x, z)\leq\deltaloc$
    \begin{align}
        \cdl||x-z||^2\leq\vdelta&(x, z)\leq\cdu||x-z||^2\\
        ||\kappa(x,z,v)-v||&\leq\kmax||x-z||\\
        \vdelta(x^+,z^+)&\leq\rho\vdelta(x, z)
    \end{align}
    with $x^+=f(x, \kappa(x, z, v))$ and $z^+=f(z, v)$ $v\in\mathcal{U}$.
\end{assumption}

	\begin{assumption}
    \label{as:localcontroller}
    There exists a local control Lyapunov-Function $\vf(x)=||x||^2{}_\mathrm{P}$ a terminal set $\xf=\{x:\vf(x)\leq\alpha_\mathrm{f}\}$ and a control law $\kappaf(x)$ such that $\forall x\in\xf$
    \begin{align}
        \label{eq:finalsetcontrol}
        f(x,\kappaf(x)) + \tilde{w} \in \xf, \forall ||\tilde{w}||\in \mathcal{W}_\mathrm{N}\\
        \label{eq:finallyap}
        \vf(f(x,\kappaf(x))) \leq \vf(x) - (||x||^2_\mathrm{Q} + ||\kappaf(x)||^2_\mathrm{R})\\
        \label{eq:finalset}
        (x, \kappaf(x)) \in \tilde{\mathcal{X}}_{N} \times \tilde{\mathcal{U}}_{N},
    \end{align}
    with $\mathcal{W}_N = \{x: \mathbb{R}^n: ||x||\leq\sqrt{\rho^N\frac{\cdu}{\cdl}}\vwmax(\hat{w})\}$.
\end{assumption}

	We use them to prove recursive feasibility.
	To prove stability, we also need the following~\cite{kohler2018novel}.
	
	\begin{assumption}
    \label{as:upboundlyap}
    There exists a $\mathcal{K}_{\infty}$-function $\gamma$ such that
\begin{equation}
    \label{eq:upboundlyap}
    \forall x\in\mathbb{R}^n: V_N(x) \leq \gamma(||x||).
\end{equation}
\end{assumption}

	
	\fakepar{Theoretical analysis}
	When we use the priority based trigger described above, the optimization problem is recursively feasible at future timesteps.

	\begin{lemma}
    \label{lem:recursivefeasibility}
    When problem (\ref{eq:optimizationproblem}) has a solution $x^*,u^*$ at $t$, then there exists a $\hat{w}_\mathrm{max}\in\mathbb{R}_{\geq0}$ such that $\forall\hat{w}\leq\hat{w}_\mathrm{max}$, at all following triggering times, i.e., at times when the central computer performs a recomputation for the corresponding system, the optimization problem has a solution.
\end{lemma}
	\begin{proof}
		We use the same techniques as the proof of~\cite[Proposition 5]{kohler2018novel}, which prooves recursive feasibility for the periodic case, and we combine them with Lemma~\ref{lem:boundedstatediff}.
		We construct a candidate solution $\tilde{x},\tilde{u}$ for the next trigger time $t_\mathrm{tr}$ with $T=t_\mathrm{tr} - t$ directly after $t$. For the next triggering timesteps, the property then follows inductively. 
		
		The candidate is based on controller $\kappa$ of Assumption~\ref{as:incrementalstability}
		\begin{align}
			\label{eq:predsystcandidate}
			\xpredcand{k+1}{t+T} =& f(\xpredcand{k}{t+T},\upredcand{k}{t+T})\nonumber\\
			\xpredcand{0}{t+T} =& \xreal{t+T}
		\end{align}
		where for $k\leq N-1-T$
		\begin{equation}
			\upredcand{k}{t+T} = \kappa(\xpredcand{k}{t+T}, \xpredopt{k+T}{t},\upredopt{k+T}{t})
		\end{equation}
		and for $k\geq N-T$
		\begin{align}
			\label{eq:predinputcandidate}
			&\upredcand{k}{t+T} =\nonumber\\ 
			&\kappa(\xpredcand{k}{t+T},\xpredopt{k+T}{t},\kappaf(\xpredopt{k+T}{t})).
		\end{align}
		We know 
		\begin{align}
			\vdelta(&\xreal{t+T}, \xpredopt{T}{t}) \nonumber\\ 
			\leq& \cdu ||\xreal{t+T} - \xpredopt{T}{t}||^2  \\
			=& \cdu ||\xreal{t+T} - \xpred{1}{t+T-1}||^2 
			\leq\cdu\vwmax(\hat{w})^2,\nonumber
		\end{align}
		because of Lemma \ref{lem:boundedstatediff}. Then, Assumption \ref{as:incrementalstability} is satisfied if $\hat{w} \leq \vwmax^{-1}(\sqrt{\frac{\deltaloc}{\cdu}})$ ($\vwmax$ is invertible as it is a $\mathcal{K}$-function (Lemma~\ref{lem:vwmaxkfunction})). It follows recursively (note $\xpredcand{0}{t+T} = \xreal{t+T}$)
		\begin{align}
			\vdelta(&\xpredcand{k}{t+T}, \xpredopt{k+T}{t}) \nonumber \\
			\leq& \rho^k \vdelta(\xpredcand{0}{t+T}, \xpredopt{T}{t})\leq\rho^k\cdu\vwmax(\hat{w})^2.
		\end{align}
		Thus
		\begin{align}
			\label{eq:statediffbound}
			||\xpredcand{k}{t+T}-&\xpredopt{k+T}{t}||^2 \nonumber \\
			\leq & \frac{1}{\cdl}\vdelta(\xpredcand{k}{t+T},\xpredopt{k+T}{t})\nonumber\\
			\leq &\rho^k\frac{\cdu}{\cdl}\vwmax(\hat{w})^2
		\end{align}
		and
		\begin{align}
			||\upredcand{k}{t+T}-&\upredopt{k+T}{t}||^2\\
			\leq& \kmax||\xpredcand{k}{t+T}-\xpredopt{k+T}{t}||^2\nonumber\\
			\leq&\rho^k\kmax^2\frac{\cdu}{\cdl}\vwmax(\hat{w})^2.\nonumber
		\end{align}
		For the state and input constraints it is for $k < N-T$ if $\hat{w}\leq\vwmax^{-1}(\frac{1}{||H||_{\infty}}\sqrt{\frac{\cdl}{\cdu}}\epsilon)$
		\begin{align}
			H&\xpredcand{k}{t+T} = H\xpredopt{k+T}{t} \nonumber \\&~~~~~~~~~~~~~~~~+ H[\xpredcand{k}{t+T} - \xpredopt{k+T}{t}]\nonumber\\
			\leq & H\xpredopt{k+T}{t} + ||H||_{\infty}||\xpredcand{k}{t+T}- \xpredopt{k+T}{t}||\cdot \mathbf{1}_\mathrm{H}\nonumber\\
			\leq & (1-\epsilon_{k+T})\cdot\mathbf{1}_\mathrm{H} + ||H||_{\infty}\sqrt{\rho^k\frac{\cdu}{ \cdl}}\vwmax(\hat{w})\cdot\mathbf{1}_\mathrm{H}\nonumber\\
			\leq & [(1-\epsilon_{k+1}) + \sqrt{\rho^k}\epsilon]\cdot \mathbf{1}_\mathrm{H}
			= (1-\epsilon_{k})\cdot\mathbf{1}_\mathrm{H} 
		\end{align}
		and for $N-1\geq k\geq N-T$ using Assumption~\ref{as:localcontroller}
		\begin{align}
			H&\xpredcand{k}{t+T} = H\xpredopt{k+T}{t} \nonumber \\&~~~~~~~~~~~~~~~~+ H[\xpredcand{k}{t+T} - \xpredopt{k+T}{t}]\nonumber\\
			\leq& H\xpredopt{k+T}{t} + ||H||_{\infty}||\xpredcand{k}{t+T} - \xpredopt{k+T}{t}||\cdot \mathbf{1}_\mathrm{H}\nonumber\\
			\leq&  (1-\epsilon_{N})\cdot\mathbf{1}_\mathrm{H} \nonumber \\ &+ ||H||_{\infty}\sqrt{\rho^k\frac{\cdu}{ \cdl}}\vwmax(\hat{w})\cdot\mathbf{1}_\mathrm{H}\nonumber\\
			\leq & [(1-\epsilon_{k+1}) + \sqrt{\rho^k}\epsilon]\cdot \mathbf{1}_\mathrm{H}
			=(1-\epsilon_{k})\cdot\mathbf{1}_\mathrm{H}
		\end{align}
		Same argumentation can be done for the input constraints, if $\hat{w} \leq \vwmax^{-1}(\frac{1}{\kmax||L||_{\infty}}\sqrt{\frac{\cdl}{\cdu}}\epsilon)$.
		
		It is $\xpredcand{N}{t+T}\in\xf$, because
		\begin{align}
			\xpredcand{N}{t+T} =& \xpredcand{N}{t+T} \nonumber \\&- \xpredopt{N+T}{t} + \xpredopt{N+T}{t} \\
			 =& \xpredcand{N}{t+T} - \xpredopt{N+T}{t} \nonumber \\ +& f(\xpredopt{N+T-1}{t}, \kappaf(\xpredopt{N+T-1}{t}))  \nonumber
		\end{align}
		and
		\begin{align}
			||\xpredcand{N}{t+T}-\xpredopt{N+T}{t}|| \leq \sqrt{\rho^N\frac{\cdu}{\cdl}}\vwmax(\hat{w}),
		\end{align} and thus $\xpredcand{N}{t+T}-\xpredopt{N+T}{t}\in\mathcal{W}_{N}$.
		
	\end{proof}

	\begin{remark}
		The bounds on $\hat{w}$ calculated are equal to the bounds in~\cite[Proposition 5]{kohler2018novel} transformed with $\vwmax(\hat{w})$. 
		Hence, if we have already designed a periodic MPC using~\cite{kohler2018novel} (plus final state constraints), then, we can reuse it for smaller disturbances in the event-triggered setting.
	\end{remark}
	\begin{remark}
		This lemma also implies that the real states and inputs always satisfy constraints $\mathcal{X}$ and $\mathcal{U}$.
	\end{remark}
	
	We now prove stability.
	First, we summarize the influence of the disturbances $w$ between $t$ and $t+T$ on the states into the artificial disturbance 
	\begin{equation}
		\wart{T}{t} 
		= \xreal{t+T}-\xpredopt{T}{t}.
	\end{equation}
	It follows from Lemma~\ref{lem:boundedstatediff} that $\wart{T}{t}$ is bounded by
	\begin{equation}
		||\wart{T}{t}|| \leq \vwmax(\hat{w}).
	\end{equation}
	
	With this, we can show the following.
	
	\begin{lemma}
		\label{lem:upperboundlyap}
		If the system (\ref{eq:system}) was recalculated at time $t$ and the next trigger time is $t+T$, then 
		there exists a $\mathcal{K}$-Function $\alpha$ such that
		for any $V_N(x(t))\leq\vwmax$
		\begin{align}
			\label{eq:upperboundlyap}
			V_N(\xreal{t+T}) \leq& V_N(x(t)) + \alpha(||\wart{T}{t}||) \\ &- \sum_{k=0}^{T-1}(||\xpredopt{k}{t}||_\rmq^2 +||\upredopt{k}{t}||_\rmr^2) \nonumber
		\end{align}
	\end{lemma}
	\begin{proof}
		In the following $\xpredcand{\cdot}{\cdot},\upredcand{\cdot}{\cdot}$ are the candidate solutions described in equations (\ref{eq:predsystcandidate})--(\ref{eq:predinputcandidate}).
		We first derive statements, which are helpful for the proof.

			\noindent
			\emph{1.} It is 
			\begin{align}
				||\xpredopt{\cdot}{t}||_\rmq^2 \leq& \frac{||Q||_2}{\lambdaminof{H^TH}}||\xpredopt{\cdot}{t}||_{H^TH}^2 \nonumber \\ 
				\leq& \frac{||Q||_2}{\lambdaminof{H^TH}}\mathbf{1}_\mathrm{H}^T\mathbf{1}_\mathrm{H} =: X_\mathrm{max}^2
			\end{align}
			\begin{align}
				||\upredopt{\cdot}{t}||_\rmr^2 \leq& \frac{||R|_2}{\lambdaminof{L^TL}}||\upredopt{\cdot}{t}||_{L^TL}^2 \nonumber \\ \leq& \frac{||R||_2}{\lambdaminof{L^TL}}\mathbf{1}_\mathrm{L}^T\mathbf{1}_\mathrm{L} =: U_\mathrm{max}^2,
			\end{align}
			where $\lambdaminof{\cdot}$ denotes the minimal eigenvalue.

			\noindent
			\emph{2.} It is (compare equation (\ref{eq:statediffbound}))
			\begin{align}
				||&\xpredcand{k}{t+T}-\xpredopt{k+T}{t}||^2 \nonumber \\ 
				\leq& \frac{1}{\cdl}\vdelta(\xpredcand{k}{t+T},\xpredopt{k+T}{t}) \nonumber \\
				\leq& \frac{1}{\cdl}\rho^k\vdelta(\xpredcand{0}{t+T},\xpredopt{T}{t}) \\
				\leq& \rho^k\frac{\cdu}{\cdl}||\xpredcand{0}{t+T}-\xpredopt{T}{t}||
				 = \rho^k\frac{\cdu}{\cdl}||\wart{T}{t}||,\nonumber
			\end{align} 
			and similar
			\begin{align}
				||\upredcand{k}{t+T}-&\upredopt{k+T}{t}||^2\nonumber \\ 
				\leq& \kmax^2 \rho^k\frac{\cdu}{\cdl}||\wart{T}{t}||.
			\end{align}
			
			\noindent
			\emph{3.} We can upper bound the cost-to-go function using equation (\ref{eq:costtogoupperbound})
			where 
			\begin{align}
				\alpha_\mathrm{1}(||\wart{T}{t}&||,k) \\
				:=& (||Q||_2+\kmax^2||R||_2)\rho^k||\wart{T}{t}||^2\nonumber\\
				&+ 2\sqrt{\frac{\cdu}{\cdl}\rho^k}||\wart{T}{t}||\nonumber\\
				&\times\left(\sqrt{||Q||_2}X_\mathrm{max}+\sqrt{\kmax^2||R||_2}U_\mathrm{max}\right)\nonumber
			\end{align}
			is a $\mathcal{K}$-function in the first argument.
			
			\newcounter{MYtempeqncnt}
			\begin{figure*}[!t]
				\vspace{0.25cm}
				\normalsize
				\setcounter{MYtempeqncnt}{\value{equation}}
				\begin{align}
					l(&\xpredcand{k}{t+T}, \upredcand{k}{t+T}) := ||\xpredcand{k}{t+T}||_\rmq^2 + ||\upredcand{k}{t+T}||_\rmr^2 \nonumber \\
					\leq&l(\xpredopt{k+T}{t}, \upredopt{k+T}{t})+||Q||_2||\xpredcand{k}{t+T}-\xpredopt{k+T}{t}||^2+||R||_2||\upredcand{k}{t+T}-\upredopt{k+T}{t}||^2\nonumber\\
					&+2\sqrt{||Q||_2}||\xpredcand{k}{t+T}-\xpredopt{k+T}{t}||||\xpredopt{k+T}{t}||_\rmq+2\sqrt{||R||_2}||\upredcand{k}{t+T}-\upredopt{k+T}{t}||||\upredopt{k+T}{t}||_\rmr\nonumber\\
					\leq&l(\xpredopt{k+T}{t}, \upredopt{k+T}{t}) + \alpha_\mathrm{1}(||\wart{T}{t}||) \label{eq:costtogoupperbound} \\[0.8em]
					||&\xpred{N}{t+T}||_\rmp^2 = ||\xpredcand{N}{t+T} - \xpredopt{N+T}{t} + \xpredopt{N+T}{t}||_\rmp^2 \nonumber\\
					\leq&||\xpredopt{N+T}{t}||_\rmp^2 + ||\xpredcand{N}{t+T} - \xpredopt{N+T}{t}||_\rmp^2 + 2||\xpredcand{N}{t+T} - \xpredopt{N+T}{t}||_\rmp||\xpredopt{N+T}{t}||_\rmp\nonumber\\
					\leq& ||\xpredopt{N+T}{t}||_\rmp^2 + ||P||_2\rho^N ||\wart{T}{t}||^2 + 2\sqrt{||P||_2\rho^N}||\wart{T}{t}||^2\alpha_\mathrm{f} \label{eq:finalstatecosupperbound}
				\end{align}

\vspace{-1.05\baselineskip}
				\hrulefill
				\vspace{-0.75cm}
				\end{figure*}

			\noindent\emph{4.} We can also upper bound the final state costs with equation (\ref{eq:finalstatecosupperbound}).
			In the following, we write $\alpha(||\wart{T}{t}||,k):=\alpha_\mathrm{1}(||\wart{T}{t}||,k)+\frac{1}{N}(||P||_2\rho^N ||\wart{T}{t}||^2 + 2\sqrt{||P||_2\rho^N}||\wart{T}{t}||^2\alpha_\mathrm{f})$, which is a $\mathcal{K}$-Function in the first argument.
			
			\noindent\emph{5.} It follows recursively from Equation (\ref{eq:finallyap})
			\begin{align}||\xpredopt{N+T}{t}&||_\rmp^2
				\leq||\xpredopt{N}{t}||_\rmp^2\\
				&- \sum_{k=N}^{N+T-1}(||\xpredopt{k}{t}||_\rmq^2 + ||\upredopt{N+n}{t}||_\rmr^2) \nonumber
			\end{align}

		\newcounter{MYtempeqncntt}
		With all of this, we can directly derive equation (\ref{eq:upperboundlyap})
		\begin{align}
			V_N(&\xreal{t+T})\label{eq:boundlyap} \\ \leq& \sum_{k=0}^{N-1}l(\xpredcand{k}{t+T},\upredcand{k}{t+T}) + ||\xpredcand{N}{t+T}||_\rmp^2\nonumber\\
			\leq& \sum_{k=T}^{N-1}l(\xpredopt{k}{t},\upredopt{k}{t}) +\sum_{k=N}^{N+T-1}l(\xpredopt{k}{t},\upredopt{k}{t}) \nonumber \\ & + ||\xpredopt{N+T}{t}||_\rmp^2 +\sum_{k=0}^{N-1}\alpha(||\wart{T}{t}||,k)\nonumber\\
			\leq& \sum_{k=0}^{N-1}l(\xpredopt{k}{t},\upredopt{k}{t}) - \sum_{k=0}^{T-1}l(\xpredopt{k}{t},\upredopt{k}{t}) \nonumber \\
			& + ||\xpredopt{N}{t}||_\rmp^2 + \sum_{k=0}^{N-1}\alpha(||\wart{T}{t}||,k)\nonumber\\
			=&V_N(\xreal{t}) - \sum_{k=0}^{T-1}l(\xpredopt{k}{t},\upredopt{k}{t}) + \alpha(||\wart{T}{t}||).\nonumber
		\end{align}
		where $\alpha(||\wart{T}{t}||) := \sum_{k=0}^{N-1}\alpha(||\wart{T}{t}||,k)$.

	\end{proof}
	
	Consider an artificial system defined as system (\ref{eq:system}) sampled at its triggering times $t_1, t_2, ...$.
	\begin{equation}
		\label{eq:artificialsystem}
		y_{n+1} = F(t_{n}, y_{n}, \wart{t_{n+1}-t_{n}}{t_{n}})
	\end{equation} 
	where $F:\mathbb{N}_0\times\mathbb{R}^n\times\mathbb{R}^n\to\mathbb{R}^n$ such that $y_{n}=x(t_n)$. 
	We now show that this system is input-to-state stable (ISS)~\cite{jiang2001input}.
	\begin{theorem}
		Consider the artificial system (\ref{eq:artificialsystem}). Assume it has infinite trigger times and problem (\ref{eq:optimizationproblem}) has a solution at $t=t_1$ and let Assumptions \ref{as:lipschitz}--\ref{as:upboundlyap} hold.
		If $\vwmax(\hat{w})\leq\hat{w}_\mathrm{max}$, then, the artificial system is ISS.
	\end{theorem} 
	\begin{proof}
		We can always find a feasible solution of the optimization problem (\ref{eq:optimizationproblem}) (Lemma~\ref{lem:recursivefeasibility}).
		Equation (\ref{eq:upboundlyap}) together with $V_N(x(t_n)) \geq ||x(t_n)||_\rmq$
		and Lemma \ref{lem:upperboundlyap} show that $V_N(y_{n})$ is an ISS-Lyapunov function for system (\ref{eq:artificialsystem}). 
	\end{proof}

	Based on this, we show stability for the entire system (\ref{eq:system}).
	\begin{theorem}
    \label{th:gISS}
    Let Assumptions \ref{as:lipschitz}--\ref{as:upboundlyap} and $\hat{w}\leq\hat{w}_\mathrm{max}$ hold. Assume that for all systems, the first optimization problem (\ref{eq:optimizationproblem}) is feasible.
    Then, the controlled system (\ref{eq:system}) is gISS.
\end{theorem}

	\begin{proof}
		Assume we have infinite trigger times.
		Because the artificial system (\ref{eq:artificialsystem}) is ISS, there exist a $\mathcal{KL}$-function $\beta$ and a $\mathcal{K}$-function $\delta$ such that~\cite{jiang2001input}
		\begin{equation}
			\label{eq:newiss}
			||y_n|| \leq \beta(||y_1||, n) + \delta(\vwmax(\hat{w})).
		\end{equation}
		Because $\gamma$ is strictly increasing 
		\begin{equation}
			V_N(y_n) \leq \gamma(||y_n||) \leq \gamma(\beta(||y_1||, n) + \delta(\vwmax(\hat{w}))).
		\end{equation}
		For every $k\in\{0,1,\cdots,t_{n+1}-t_n\}$ it is
		\begin{equation}
			V_N(\xpred{k}{t_n}) \leq V_N(\xpred{0}{t_n}) = V_N(y_n),
		\end{equation}
		which follows by applying equation (\ref{eq:boundlyap}) without disturbances. 
		Further, it is
		\begin{align}
			\lambdaminof{Q}&||\xpred{k}{t_n}|| \leq ||\xpred{k}{t_n}||_\rmq \leq  V_N(\xpred{k}{t_n})
		\end{align}
		and for $t\leq t_{1}$ (recall $u(t<t_{1}) = 0$ and $t_1\leq p-1$)
		\begin{align}
			\label{eq:beginningupperbound}
			||\xreal{t}|| \leq \vw(\hat{w}, ||\xreal{0}||, t) \leq \vw(\hat{w}, ||\xreal{0}||, p-1)
		\end{align}

		Applying Lemma \ref{lem:boundedstatediff} then gives with $t=k+t_n$, $k\geq0$
		\begin{align} 
		||\xreal{t}|| \leq& ||\xpred{k}{t_n}|| +\vwmax(\hat{w}) \\
		 \leq& \frac{1}{\lambdaminof{Q}}\gamma(\beta(||y_1||, n) + \delta(\vwmax(\hat{w}))) + \vwmax(\hat{w})\nonumber\\
		 \leq& \frac{1}{\lambdaminof{Q}}\gamma(\beta(\vw(\hat{w}, ||\xreal{0}||, p-1), n)\nonumber\\
		&+ \delta(\vwmax(\hat{w}))) + \vwmax(\hat{w}).\nonumber
		\end{align}	

		For $t \leq t_1$, we can upper $||\xreal{t}||$ by Equation (\ref{eq:beginningupperbound}).

		If we have a finite amount of trigger times, we can apply the same arguments as before until the last trigger time. 
		After this, we know, the predicted state converges to zero, and the real state has a difference of at most $\vwmax(\hat{w})$.
	\end{proof}

	\section{EXPERIMENTS}

	To demonstrate our approach, we consider position control of quadcopters in simulation.
	We use a standard quadcopter kinetic model with 13 states (position, velocity, quaternions, angular velocity).
	We disturb the velocities with uniformly distributed noise.
	The quadcopters need to fly to a target position and stay there.
	They fly in distinct spaces, i.e., we do not consider collision avoidance.
	Our MPC uses a horizon of $T=25$ and a sampling time of $0.1\si{\second}$. 
	The exact hyperparameters together with the entire code can be found in
	\url{https://github.com/Data-Science-in-Mechanical-Engineering/event-triggered-robust-mpc}.

\begin{figure}[t]
		\centering
		\newcommand{\plotsep}{0.2\linewidth}
		\newcommand{\plotheight}{0.32\linewidth}
		\newcommand{\plotwidth}{0.9\linewidth}
		\newcommand{\plotmarkersize}{1pt}
		\vspace{0.2cm}
		\begin{tikzpicture}
			\coordinate (cx) at (0.0, 0.0);
			\coordinate [below=\plotsep of cx] (cy);
			\coordinate [below=\plotsep of cy] (cz);
			\coordinate [below=\plotsep of cz] (cprio);
			\coordinate [below=\plotsep of cprio] (ctriggered);

			\begin{axis}
				[at=(cx),
				anchor=north,
				width=\plotwidth,
				height=\plotheight,
				ylabel={$x (\si{\meter})$},
				xmin=-0.01, xmax=3.91,
				]
			\addplot+ [mark=none, solid] table [x=time,y=quad_0_0, col sep=comma] {images/example_run.csv};
			\addplot+ [mark=none, solid] table [x=time,y=quad_1_0, col sep=comma] {images/example_run.csv};    
			\addplot+ [mark=none, solid] table [x=time,y=quad_2_0, col sep=comma] {images/example_run.csv};    
			\end{axis}

			\begin{axis}
				[at=(cy),
				anchor=north,
				ymin=-0.4,
				width=\plotwidth,
				height=\plotheight,
				ylabel={$y (\si{\meter})$},
				xmin=-0.01, xmax=3.91,
				]
			\addplot+ [mark=none, solid] table [x=time,y=quad_0_1, col sep=comma] {images/example_run.csv};
			\addplot+ [mark=none, solid] table [x=time,y=quad_1_1, col sep=comma] {images/example_run.csv};    
			\addplot+ [mark=none, solid] table [x=time,y=quad_2_1, col sep=comma] {images/example_run.csv};    
			\end{axis}

			\begin{axis}
				[
				at=(cz),
				anchor=north,
				width=\plotwidth,
				height=\plotheight,
				ylabel={$z (\si{\meter})$},
				xmin=-0.01, xmax=3.91,
				]
			\addplot+ [mark=none, solid] table [x=time,y=quad_0_2, col sep=comma] {images/example_run.csv};
			\addplot+ [mark=none, solid] table [x=time,y=quad_1_2, col sep=comma] {images/example_run.csv};    
			\addplot+ [mark=none, solid] table [x=time,y=quad_2_2, col sep=comma] {images/example_run.csv};    
			\end{axis}


			\begin{axis}[at=(cprio),
				anchor=north,
				xlabel={Time(s)}, 
				xlabel style={yshift=2mm},
				ylabel style={align=center}, ylabel={Selected\\quadcopter},
				ymin=0.9,ymax=3.01,
				ytick={1, 2, 3},
				xmin=-0.001, xmax=3.91,
				height=\plotheight, 
				width=\plotwidth,
					grid style={black!50},
					grid=both, ymajorgrids=false, yminorgrids=true, xmajorgrids=false,
					]
					
					\addplot [color=black, only marks, mark=square*, mark size=0.9pt] table [x=time,y=selected_0, col sep=comma]  {images/example_run.csv};
				

		
				\end{axis}
		\end{tikzpicture}
		\vspace{-0.3cm}
		\caption{Example trajectory flown by three quadcopters, while the central computer controls only one of them at a time.}
		\vspace{-0.4cm}
		\label{fig:exampletrajectory}
	\end{figure}
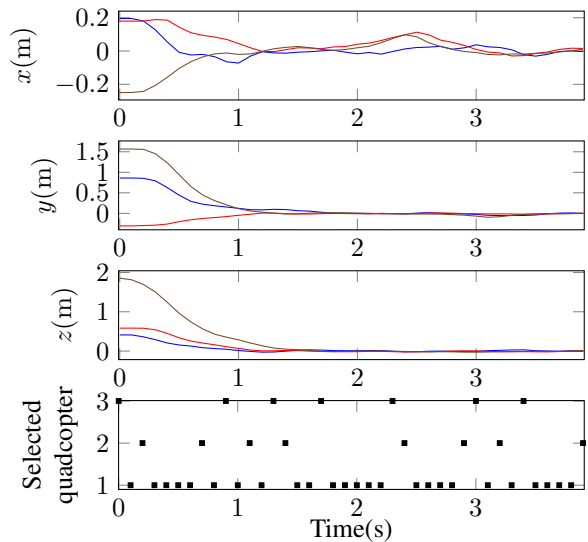

	Figure~\ref{fig:exampletrajectory} visualizes the behavior of our ET-MPC controlling $M_\mathrm{s}=3$ quadcopters with $M_\mathrm{c}=1$. 
	We set the first quadcopter's disturbance to have twice the amplitude than the other quadcopters.
	The ET-MPC stabilizes all quadcopters despite disturbances while selecting only one quadcopter for computation at a time.
	The first quadcopter is selected more often for computation, showing the effect of the ET distributing resources to systems needing them the most.

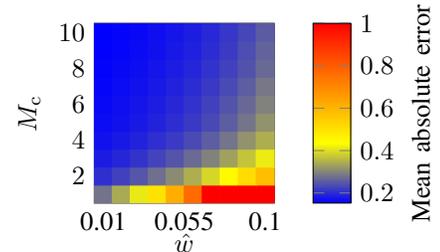
\begin{figure}[t]
		\centering
		\begin{tikzpicture}
			\begin{axis}
				[colorbar,
				colormap/hot,
				colorbar style={
				ylabel=Mean absolute error,
				ylabel style={yshift=0cm},
				},
				xmin=0.5,xmax=10.5,
				ymin=0.5,ymax=10.5,
				width=0.46\linewidth,
				height=0.46\linewidth,
				xlabel={$\hat{w}$},
				xlabel style={yshift=2mm},
				ylabel={$M_\mathrm{c}$},
				zlabel ={a},
				xtick={1, 5.5, 10},
				xticklabels = {0.01, 0.055, 0.1},
				]
			\addplot [matrix plot*, 
			mark=none,
		mesh/cols=10,
		point meta=explicit] table [x=i,y=j,meta=result, col sep=comma] {images/param_sweep.csv};    
			\end{axis}
		\end{tikzpicture}
		\vspace{-0.35cm}
		\caption{Behavior of ten controlled quadcopters for different amplitudes of disturbances and computation resources. \capt{The plot shows the mean absolute deviation of the position from 0. Areas marked as red (value 1) are unstable.}}
		\vspace{-0.5cm}
		\label{fig:parametersweep}
	\end{figure}

	We further investigate the behavior of our controller for larger setups in Fig.~\ref{fig:parametersweep}.
	We vary $M_\mathrm{c}$ and the maximum disturbance $\hat{w}$.
	For every combination of $M_\mathrm{c}$ and $\hat{w}$, we run 20 simulations with randomly sampled initial positions.
	If $\hat{w}$ is too large and $M_\mathrm{c}$ is too small, the quadcopters are unstable (cf. Theorem~\ref{th:gISS}).
	The higher $M_\mathrm{c}$, i.e., the more compute power, the better the performance, especially for larger disturbances.
	The same holds true for lower $\hat{w}$.
	However, one can see a plateau forming for higher $M_\mathrm{c}$ and lower disturbances. 
	Increasing the central computer's power hence does not change the performance significantly at some point.
	Our approach allows to save computational resources without sacrificing performance.

	\section{CONCLUSION}
	We presented a novel event-triggered robust MPC, which takes strictly limited resources of central computer servers into account, by only computing a new control input for a subset of systems.
	Through a theoretical analysis, we showed that our approach leads to stable closed loop systems.
	In experiments, we demonstrated that our approach is able to control many systems with limited resources.

	The current approach is limited to independent systems.
	In future work, we will investigate how to extend the approach to interconnected systems.

	\section{ACKNOWLEDGMENTS}

	We thank Henrik Hose, David Stenger and Christian Fiedler for helpful discussions.

	\bibliographystyle{IEEEtran}
	\bibliography{references}

\end{document}